\documentclass[%
 reprint,
 amsmath,amssymb,
 aps,
]{revtex4-2}
\usepackage{xcolor}
\usepackage{graphicx}
\usepackage{dcolumn}
\usepackage{bm}

\begin{document}

\preprint{APS/123-QED}

\title{Generalized epidemic model incorporating non-Markovian infection processes and waning immunity}

\author{Qihui Yang$^1$$^,$$^3$}
\thanks{qihui@ksu.edu.}

\author{Joan Saldaña$^2$$^,$$^3$}
\thanks{joan.saldana@udg.edu}
\author{Caterina Scoglio$^1$}
\thanks{caterina@ksu.edu}

\affiliation{
$^1$Department of Electrical and Computer Engineering, Kansas State University, Manhattan 66506, Kansas, USA}

\affiliation{%
 $^2$Department of Computer Science, Applied Mathematics, and Statistics, Universitat de Girona, Girona 17003, Catalonia, Spain}%
 
 \affiliation{%
 $^3$These authors contributed equally to this work.
 }

\date{\today}

\begin{abstract}
The Markovian approach, which assumes exponentially distributed interinfection times, is dominant in epidemic modeling. However, this assumption is unrealistic as an individual’s infectiousness depends on its viral load and varies over time. In this paper, we present a Susceptible-Infected-Recovered-Vaccinated-Susceptible epidemic model incorporating non-Markovian infection processes. The model can be easily adapted to accurately capture the generation time distributions of emerging infectious diseases, which is essential for accurate epidemic prediction. We observe noticeable variations in the transient behavior under different infectiousness profiles and the same basic reproduction number $\mathcal{R}_0$. The theoretical analyses show that only $\mathcal{R}_0$ and the mean immunity period of the vaccinated individuals have an impact on the critical vaccination rate needed to achieve herd immunity. A vaccination level at the critical vaccination rate can ensure a very low incidence among the population in the case of future epidemics, regardless of the infectiousness profiles.
\end{abstract}

\maketitle

\section{Introduction}
\label{sec:level1}

The widely used formulation of compartmental epidemic models in terms of ordinary differential equations (ODEs)  implicitly assumes both a constant probability per unit of time of leaving the infectious state (recovery rate) and a constant transmission probability per unit of time (transmission rate). This is analogous to the setting where the sojourn times in the infectious state (infectious period) and the generation (or interinfection) times are exponentially distributed. Following \cite{favero2022modelling, Wallinga2007}, we define generation times as the time between the infection of a secondary case and the infection of the corresponding primary case.

However, many empirical studies have shown that the exponential distribution does not fit well clinical data about sojourn times in several compartments of an infectious disease model. For example, several studies have shown that the generation times for the spreading of severe acute respiratory syndrome coronavirus 2 (SARS-CoV-2) are not exponentially distributed \citep{ferretti2020,li2020early,nishiura2020serial}. This necessitates the development of proper epidemic models that consider nonexponential sojourn times.

Already in the foundational paper by \cite{kermack1927contribution}, the formulation of the Susceptible-Infected-Recovered (SIR) model assumed a transmission probability depending on the time after infection, also called the age of infection. The reason for that is pretty clear: an individual's infectiousness depends on their viral load, which, in turn, varies over time. Similar ages are also introduced when the probability of processes like loss of immunity depends on the time after entering the recovered state (time since clearance). In such cases, the dynamics are described by non-Markovian processes, as the current status of individuals depends on their complete history within a given compartment. Consequently, the sojourn time in each state and the generation time no longer follow an exponential distribution.

In a deterministic context, this fact leads to the formulation of epidemic models in terms of partial differential equations (PDEs), where a population is described by densities with respect to one or more of those times or ages. For the age-of-infection SIR model, the PDE corresponds to the so-called McKendrick-von Foester equation (see \cite{reyne2022} for a model with several ages, including the age of vaccination). Such a formulation, equivalent to renewal equations under enough regularity conditions \cite{iannelli2017basic}, allows the analysis of the impact of non-Markovian processes on the epidemic spread. Recently, a PDE formulation at the node level has also been used to model the epidemic spread on complex networks \cite{feng2019equivalence, han2023non}.  

Staged-progression epidemic models are an alternative way to model non-Markovian epidemics. These models are halfway between simple ODE compartmental models and PDE models because they consider a sequence of different lengths of infectious stages (compartments). Each of them has its own recovery rate and transmission rate \citep{Brauer2008, Hyman1999}. So, they can be considered as a sort of discretization of the PDE models \citep{Landry2021}. Indeed, these models have been used in the literature to approximate nonexponential infectious periods by subdividing the infectious compartment into several subcompartments with exponentially distributed infectious periods. The original distribution is then approximated by a sum of exponential distributions \citep{Roberts2007}. 

In this paper, we formulate a Susceptible-Infected-Recovered-Vaccinated-Susceptible (SIRVS) epidemic model and provide theoretical analyses of the model regarding the equilibria and the critical vaccination rate. Following \cite{Anderson2020}, the latter is obtained from the bifurcation from the disease-free equilibrium where susceptible and vaccinated individuals are present. We perform PDE numerical integration and agent-based simulations to examine the impact of infectiousness profiles and vaccination rates on epidemic dynamics under these two approaches. In particular, agent-based simulations allow to assess the impact of population sizes on the occurrence of secondary waves.

The contributions of the paper are summarized as follows: 

$\bullet$ We present a general method to  model non-Markovian infection processes from rate-based transitions. 
In the agent-based simulations, transitioning from Markovian infection processes to non-Markovian infection processes is achieved by adjusting the value of infectiousness parameter, which results in the desired generation time distributions. This implementation option provides a straightforward way to create comparable agent-based models from PDE models. 

$\bullet$ We include the effects of recovery while calculating infectiousness profiles, which is more realistic compared to previous models which model the infectiousness profiles independent of the recovery.

$\bullet$ With the same $\mathcal{R}_0$, we observe significant differences in the transient phase between non-Markovian and Markovian models, and the magnitude of the differences is affected by the infectious period. The transient phase refers to the early stages of the epidemic dynamics, when the number of infections changes and the system is far from the steady state.

$\bullet$ We provide equilibrium analyses of the model and conclude that only $\mathcal{R}_0$ and the mean immunity period of the vaccinated individuals have an impact on the critical vaccination rate needed to achieve herd immunity.

$\bullet$ A continuous vaccination of the population at the predicted critical rate ensures a very low incidence among the population in case of future epidemics, regardless of the infectiousness profiles. 

$\bullet$ To the best of the authors' knowledge, the work for the first time, explores the potential contribution of agent-based models contrasted with PDE models in non-Markovian epidemic modeling. We observe the median values of simulation results with secondary waves are close to the results of the deterministic PDE model for population sizes sufficiently large. In contrast, simulations produce patterns not predicted by the PDE model when population sizes are sufficiently small and the stochastic extinction of the disease becomes an important factor after an initial outbreak. 

\section{\label{sec:section22} The reproduction number and generation times}
Suppose the recovery rate $\gamma$ and the \textit{infectiousness} (per-contact transmission probability) $\beta$ are both functions of the age of infection $\tau$, i.e., $\gamma = \gamma(\tau)$ and $\beta = \beta(\tau)$, and we assume a constant contact rate $c$ per individual in a randomly mixed population. In that case, the basic reproductive number $\mathcal{R}_0$ is the sum of the infections caused by an infected individual at each age of infection in a totally susceptible population, conditioning on the probability of being infectious at each age. So, we have  
\begin{equation}
\mathcal{R}_0 = c \int_0^\infty \beta(\tau) \, e^{-\int_0^\tau \gamma(s) \, ds}  \, d\tau,
\label{R0}
\end{equation}
where the exponential term is the probability of being infectious at time $\tau$ since infection, and $\eta(\tau) = c \, \beta(\tau) \, e^{-\int_0^\tau \gamma(s) \, ds}$ is the \textit{infectivity} of an individual at the age of infection $\tau$ \citep{Roberts2007}. In other models, the contact rate $c$ is included in the definition of $\beta$, which is then called \textit{effective contact rate} \citep{Vynn2010}. 

A simple but important remark follows from \eqref{R0}, namely, if $\beta$ is constant, then $R_0$ depends on the mean infectious period $\bar{\tau}_I$ but not on its particular distribution: $\mathcal{R}_0 = c \, \beta \, \bar{\tau}_I$. A similar result follows for staged-progression models if $\beta$ is constant in each compartment: $\mathcal{R}_0 = \sum_{i=1}^{n_{ic}} \mathcal{R}_0^i = c \sum_{i=1}^{n_{ic}} \beta_i \bar{\tau}_I^i$ with $n_{ic}$ being the number of infectious compartments \citep{Brauer2008}. 

Normalizing the infectivity $\eta(\tau)$ by $\mathcal{R}_0$ we obtain the probability density function (PDF) of the generation times during the initial phase of an epidemic \citep{Ma2020,Wallinga2007}: 
\begin{equation}
w(\tau) = \frac{\eta(\tau)}{\mathcal{R}_0} = \frac{c \, \beta(\tau) e^{-\int^{\tau}_0\gamma(s)\,ds}}{\mathcal{R}_0}. 
\label{w_beta}
\end{equation}

The interinfection times generated according to this time-independent PDF have been called \textit{intrinsic} generation times to distinguish them from the \textit{realized} generation times as the epidemic progresses \citep{Champredon}. The realized generation time distribution changes over time due to changes in individuals' contact patterns, the depletion of the susceptible population, and the competition among infectors \cite{Champredon,Hart2022,Nishiura,Torneri}. 

 From the relationship between the transmission probability $\beta(\tau)$, the recovery rate $\gamma(\tau)$, and the generation time distribution $w(\tau)$ given by Eq~\eqref{w_beta}, it follows that an equivalent approach to study non-Markovian infection processes is the one based on the distribution itself of the generation times during the epidemic spread. For instance, such an approach has been used to simulate stochastic epidemics on networks. This relationship clearly shows that changing the profile of $w(\tau)$ will affect the epidemic threshold because it implies a change in the profile of $\beta(\tau)$, even though the mean infectious period and the mean transmission rate are kept the same. This is what was observed in \cite{van2013non}. 

The empirical knowledge of $w(\tau)$ at the beginning of an epidemic helps to estimate $\mathcal{R}_0$ from the initial epidemic growth rate $r$ by means of the relation \citep{Ma2020,Roberts2007,Wallinga2007}:
\begin{equation}
   \int_0^\infty e^{-r\tau} w(\tau) d\,\tau = \frac{1}{\mathcal{R}_0},
\label{Euler_Lotka}
\end{equation}
obtained from the Euler-Lotka equation after replacing $\eta(\tau)$ by $\mathcal{R}_0 \, w(\tau)$. This expression also says that if we set $\mathcal{R}_0$ to a fixed value, then different generation time distributions $w(\tau)$ will lead to different initial epidemic growth rates $r$ and, hence, to different transient behaviors of the epidemic. 

\section{\label{sec:section2}The SIRVS model}
In this paper, we generalize the SIRVS epidemic model with waning immunity for recovered (R) and vaccinated (V) individuals considered in \cite{Anderson2020} by introducing an age of infection for the individuals in the I compartment, and an age of immunity (time since clearance) for individuals in the R and V compartments. 

The population in each compartment at time $t$ is then described by the densities $I(t,\tau)$, $R(t,\tau)$ and $V(t,\tau)$ with respect to the corresponding sojourn time $\tau$ in the compartment. As in \cite{Anderson2020}, the epidemic time scale is supposed to be much faster than the time scale for demographic processes (growth, births, and deaths), which allows us to consider that the population remains constant and equal to $N$, that is, 
$$
S(t) + \int_0^\infty I(t,\tau)d\tau + \int_0^\infty R(t,\tau)d\tau + \int_0^\infty V(t,\tau)d\tau = N.
$$
Moreover, we assume that the recovery rate $\gamma(\tau)$ satisfies that $\lim\limits_{\tau \to \infty} \left(\tau \, e^{-\int_0^{\tau} \gamma(s) \,ds}\right)=0$. The same condition is satisfied by the  rates $\delta(\tau)$ and $\delta^v(\tau)$ of immunity loss in the R and V compartments, respectively. This hypothesis guarantees a finite mean sojourn time $\overline{\tau}$ in any of these compartments:
$$
\overline{\tau}_\alpha = \int_0^\infty \tau \alpha(\tau) e^{-\int_0^{\tau} \alpha(s)\,ds} d\tau = \int_0^{\infty}  e^{-\int_0^{\tau} \alpha(s)\,ds} d\tau < \infty,
$$
with $\alpha(\tau)=\gamma(\tau), \, \delta^v(\tau), \, \delta(\tau)+v$. Here, $v \ge 0$ stands for the vaccination rate of susceptible and recovered individuals.   

According to the previous assumptions, the equations governing the dynamics of the SIRVS model are given by
\begin{align*}
\frac{\partial I}{\partial t} + \frac{\partial I}{\partial \tau} & =  -\gamma(\tau) I(t,\tau), 
\\
\frac{\partial V}{\partial t} + \frac{\partial V}{\partial \tau} & =  -\delta^v(\tau) V(t,\tau),   
\\
\frac{\partial R}{\partial t} + \frac{\partial R}{\partial \tau} & =  -(\delta(\tau)+v) R(t,\tau), 
\\
\frac{d S}{dt} & =  \int_0^\infty \left[  \delta^v(\tau) V(t,\tau) + \delta(\tau) R(t,\tau) \right] d\tau
\\
& \quad - S \phi(t) - v S,
\end{align*}

where $\phi$ denotes the force of infection (the rate at which a susceptible individual becomes infected) and is given by

$$
\phi(t) = c \int_0^\infty \beta(\tau) \frac{I(t,\tau)}{N} \, d\tau. 
$$

These equations are endowed with the boundary conditions at $\tau=0$

\begin{align*}
I(t,0) & = S(t)\phi(t), 
\\
V(t,0) & = v S(t) + v \int_0^\infty R(t,\tau)\,d\tau, 
\\
R(t,0) & = \int_0^\infty  \gamma(\tau) I(t,\tau) d\tau, 
\end{align*} 
and the initial condition $I(0,\tau) = I^0(\tau)$, $V(0,\tau) = V^0(\tau)$, 
$R(0,\tau) = R^0(\tau)$, and $S(0)=S^0$. 

Note that, if all the rates are constant, we obtain the original ODE model in \cite{Anderson2020} by integrating the first three equations of the SIRVS model with respect to $\tau$.

\section{Equilibria and the critical vaccination rate}{\label{sec:level3}}

Using that $\int_0^\infty  \gamma(\tau) \, e^{-\int_0^\tau \gamma(s) \, ds} d\tau = 1$ and $\lim\limits_{\tau \to \infty} \left(\tau e^{-\int^{\tau}_0 (\delta(s)+v) \,ds} \right) = 0$, it follows that the equilibrium densities satisfy 
\begin{align*}
I^*(\tau) & = S^* \phi^* e^{-\int_0^\tau \gamma(s) \, ds}, \\
V^*(\tau) & = v S^* \left( 1 + \phi^* \, \overline{\tau}_{\tilde \delta} \right) e^{-\int_0^\tau \delta^v(s) \, ds}, \\
R^*(\tau) & = S^* \phi^* e^{-\int_0^\tau (\delta(s) + v) \, ds},
\end{align*}
where $\displaystyle \phi^* = \frac{c}{N} \int_0^\infty \beta(\tau) I^*(\tau) d\tau$ is the equilibrium force of infection, and $\overline{\tau}_{\tilde \delta} = \int_0^\infty e^{-\int_0^\tau (\delta(s)+v)ds} d\tau$ is the mean sojourn time in the R compartment. Note that $\overline{\tau}_{\tilde \delta}$ takes into account that an R individual can lose its immunity and become susceptible or, alternatively, it can move to the V compartment if vaccinated. 

Introducing the expression of $I^*(\tau)$ into that of $\phi^*$ and using \eqref{R0}, it follows
$$
\phi^* =  \phi^* \frac{S^*}{N} \mathcal{R}_0.
$$
So, either $\phi^*=0$, which corresponds to the disease-free equilibrium (DFE), or $\phi^*>0$ and then $\mathcal{R}_0 S^*/N=1$, which corresponds to the unique endemic equilibrium. 

The DFE is then given by $I^*(\tau)=0$, $R^*(\tau)=0$, and
$$
S^* = \frac{N}{1+v\,\overline{\tau}_{\delta^v}}, \quad V^*(\tau) =  \frac{v \, N}{1+v\,\overline{\tau}_{\delta^v}} e^{-\int_0^{\tau} \delta^v(s)\,ds}, 
$$
where $\overline{\tau}_{\delta^v}$ is the mean immunity period of vaccinated individuals and it is used that $S^*+\int_0^\infty V^*(\tau)\,d\tau=N$. As expected, if $v=0$ then $S^*=N$.

At the endemic equilibrium $(I^*(\tau), V^*(\tau), R^*(\tau))$, the fraction of susceptible individuals at equilibrium is
$$
s^* = \frac{S^*}{N} = \frac{1}{\mathcal{R}_0},
\label{condS*}
$$
which is the same well-known relationship between $s^*$ and  $\mathcal{R}_0$ as for the standard SIS (and SIRS) models (\cite{Keeling}). Note that, to have an endemic equilibrium ($s^*<1$), $\mathcal{R}_0 > 1$. 

The value of $\phi^*$ is obtained from the condition
$$
S^* + \int_0^{\infty} I^*(\tau)\,d\tau + \int_0^{\infty} V^*(\tau)\,d\tau 
+ \int_0^{\infty} R^*(\tau)\,d\tau = N,  
$$
which amounts to
$$
\overline{\tau}_\gamma \phi^* + v\, \overline{\tau}_{\delta^v} + v \, \overline{\tau}_{\delta^v} \overline{\tau}_{\tilde \delta} \, \phi^* + \overline{\tau}_{\tilde \delta} \, \phi^* = \mathcal{R}_0-1,
$$
where $\overline{\tau}_\gamma$ is the mean infectious period. So, the force of infection at the endemic equilibrium is given by
\begin{equation}
\phi^* = \frac{\mathcal{R}_0 - 1 -v \, \overline{\tau}_{\delta^v}}{\overline{\tau}_\gamma + \overline{\tau}_{\tilde \delta}(1 + v \, \overline{\tau}_{\delta^v})}.
\label{phi*}
\end{equation}
Note that, with vaccination, $\mathcal{R}_0 > 1$ does not guarantee $\phi^* > 0$. Now, it is needed that $R_0 > 1 + v \, \overline{\tau}_{\delta^v}$.

So, assuming this condition and dividing the equilibrium densities by the total population $N$, it follows that the normalized equilibrium densities $i^*$, $w^*$ and $r^*$ in the I, V and R compartments, respectively, are given by 
\begin{align*}
i^*(\tau) & =  \frac{1}{\mathcal{R}_0}   \frac{\mathcal{R}_0 - 1 - v \, \overline{\tau}_{\delta^v}}{\overline{\tau}_\gamma + \overline{\tau}_{\tilde \delta} (1 + v \, \overline{\tau}_{\delta^v})} \, e^{-\int_0^\tau \gamma(s)\,ds}, \\
w^*(\tau) & = \frac{v}{\mathcal{R}_0} \left( 1 + \overline{\tau}_{\tilde \delta} \, \frac{\mathcal{R}_0 - 1 - v \, \overline{\tau}_{\delta^v}}{\overline{\tau}_\gamma + \overline{\tau}_{\tilde \delta} (1 + v  \, \overline{\tau}_{\delta^v})} \right) \, e^{-\int_0^\tau \delta^v(s)\,ds}, \\
r^*(\tau) & = \frac{1}{\mathcal{R}_0}   \frac{\mathcal{R}_0 - 1 - v \, \overline{\tau}_{\delta^v}}{\overline{\tau}_\gamma + \overline{\tau}_{\tilde \delta} (1 + v \, \overline{\tau}_{\delta^v})} \, e^{-\int_0^\tau (\delta(s)+v)\,ds}
\end{align*}

The condition for a  bifurcation from the DFE is obtained by imposing that the right-hand side of \eqref{phi*} is equal to 0. In particular, using $v$ as a tuning parameter, the resulting critical vaccination rate is 
$$
v_c = \frac{\mathcal{R}_0-1}{\overline{\tau}_{\delta^v}}.
$$
 Note that, since this paper considers waning immunity, continuous vaccination campaigns are required to preserve herd immunity. The critical vaccination rate defines the minimum supply of vaccine that ensures that the system always reaches the DFE after the introduction of new cases. In other words, this vaccination rate confers herd immunity to the population and thus prevents future major epidemic outbreaks. Interestingly, only the mean immunity period of vaccinated individuals (but not the distribution of its duration) is relevant for $v_c$. In particular, the threshold condition obtained in \cite{Anderson2020} follows from this expression after replacing $\overline{\tau}_{\delta^v}$ by $1/\delta_0^v$, the mean duration of immunity arising from vaccination when $\delta^v$ is constant and equal to $\delta_0^v$.   

\section{Agent-based stochastic simulations}
\begin{figure}[ht]
\centering
\includegraphics[width=65mm]{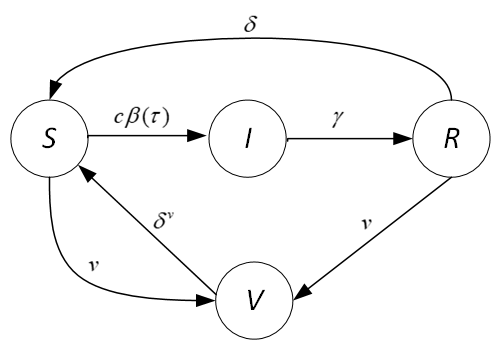}
\caption{\label{fig:statechart} SIRVS model transition process. Each circle represents one of the disease states: susceptible (S), infected (I), recovered (R), and vaccinated (V). Symbols above the arrows indicate the rates of transitions between the states.}
\end{figure}

To perform stochastic simulations, we reconceptualize the mathematical formulations from an agent-based perspective. The PDE models adopt an aggregate representation of the entire population. In comparison, agent-based models (ABMs) enable us to analyze the overall system behavior emerging from autonomous agents’ behaviors and interactions. In the model, each person agent follows the SIRVS transition process as shown in Fig~\ref{fig:statechart}. 

At the start of the simulation, all individuals are equally susceptible, except a small fraction of the population that will be randomly selected to enter the infectious state to start the epidemic. Each person agent $i$ records the time when it transitioned to the infectious state (becomes infected), denoted by $t_0(i)$. Each person contacts $c$ number of other agents on average per day. Every time an infectious person agent $i$ executes a contact event, it will fire an infection event based on a probability $\beta(\tau)$, where $\tau$ is the age of infection (current time $t-t_0 (i)$). If the infection event happens, person-agent $i$ will randomly select a person-agent $j$ from the whole population to transmit the infection. If the selected person $j$ is in the susceptible state, person $j$ will immediately transition to the infectious state. Otherwise, person $j$ will remain in its current state. Later, person $i$ will leave the infected state and transition to the recovered state at recovery rate  $\gamma(\tau)=1/\bar \tau_\gamma$, leading to exponentially distributed infectious periods. In addition, a person agent in the recovered state or susceptible state will transition to the vaccinated state according to the same rate $v$ as defined by the PDE model. As immunity wanes over time, a person in the vaccinated state or in the recovered one will transition to the susceptible state  based on an immunity loss rate $\delta=\delta^v$. 

\section{Results}
In this section, we compare the epidemic dynamics of the SIRVS models with different infectiousness profiles, infectious periods, and vaccination rates. 

\subsection{General setup}
Results are obtained from both agent-based simulations and the PDE model formulation. The ABMs are implemented in the AnyLogic 8 university researcher version. The PDE system of the model is numerically integrated by using a finite difference scheme based on the one introduced in \cite{ ackleh2005}. Agent-based simulations are performed with the same parameters values as the mathematical model. There are 500 simulation runs for each scenario. In all scenarios, the basic reproduction number $\mathcal{R}_0$ is set to be 2.5, and the contact rate $c$ equals 10. In addition, we assume recovered and vaccinated individuals have the perfect protection against infections for a mean immunity period of six months based on references \citep{ukreport,dan2021immunological}.
In particular, the rates of immunity loss are assumed to be constant and equal to $\delta=\delta^v=0.0055$. For these values of $\mathcal{R}_0$ and $\delta^v$, the corresponding critical vaccination rate is $v_c=0.00825$. Finally, the recovery rate $\gamma(\tau)$ is also assumed to be constant and, hence, equal to $1/\bar \tau_\gamma$. 

 For scenarios with constant infectiousness, $\beta(\tau)$ is constant and equal to $\mathcal{R}_0 \gamma /c$. In this case, since $\gamma$ is also constant, the generation time is exponentially distributed: $w(\tau) = \gamma e^{-\gamma \tau}$ (cf. with Eq.~\eqref{w_beta}). For varying infectiousness profiles, following \cite{ferretti2020}, we assume that $w(\tau)$ follows a Weibull distribution, i.e., $w(\tau) =\frac{k}{\lambda}(\frac{\tau}{\lambda})^{k-1}e^{-(\tau/\lambda)^k}$, with the shape parameter $k$ kept unchanged and changing the scale parameter $\lambda$ to result in the desired value of the mean generation time (MGT). More precisely, from the expression of the MGT of a Weibull distribution, it follows that $\lambda = \frac{\mbox{MGT}}{\Gamma \left(1+\frac{1}{k} \right)}$, where $\Gamma$ denotes the gamma function. Then, $\beta(\tau)$ will follow from Eq.~\eqref{w_beta} with $\mathcal{R}_0=2.5$ and $c=10$. So, the generation time distribution $w(\tau)$ is introduced only to obtain $\beta(\tau)$, which is then used to trigger an infection event once an infectious contact has occurred. In other words, simulations did not use timeout-triggered infection transmissions based on $w(\tau)$, but rather rate-based transitions (see Fig.~\ref{fig:statechart}). The realized generation times are then recorded to check the accuracy of the procedure. 

\begin{figure}[htbp]
\centering
\includegraphics[width=90mm]{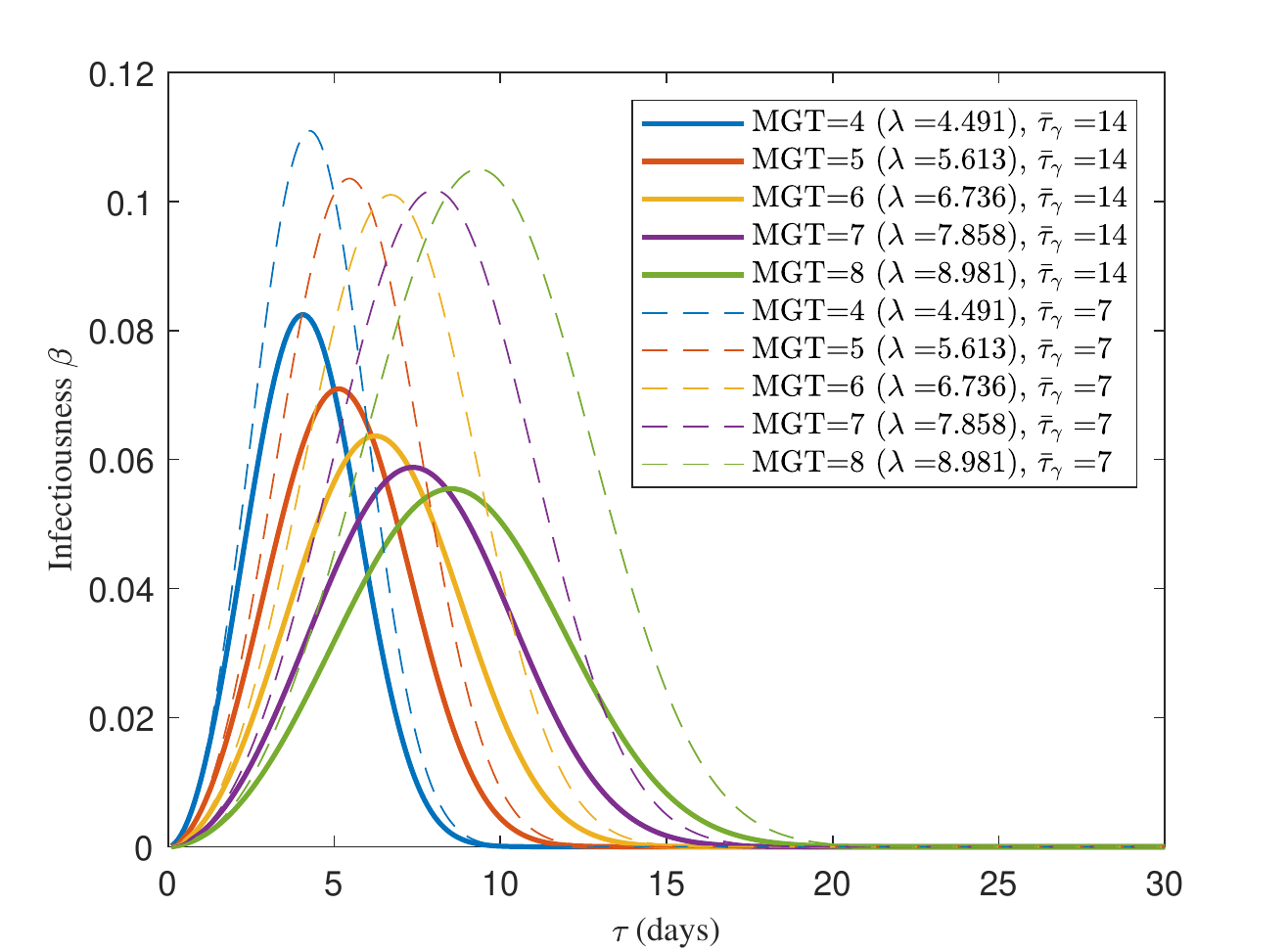}
\caption{\label{fig:iprofile} Infectiousness profiles. The solid lines correspond to five infectiousness profiles associated with a mean infectious period $\bar{\tau}_\gamma=14$ days to obtain the same $\mathcal{R}_0=2.5$. The dashed lines show five infectiousness profiles associated with a mean infectious period $\bar{\tau}_\gamma=7$ days to achieve the same $\mathcal{R}_0=2.5$. Colors (blue, orange, yellow, purple, green) correspond to the five mean values for the generation times, varying between 4 and 8 days with an interval of 1 day. The Weibull shape parameter $k=2.826$ for all curves.} 
\end{figure}

There exists a variety of estimated epidemiological parameters for COVID-19. For example, the MGT of the alpha and delta SARS-CoV-2 variants are estimated between 3.44 and 7.5 days \citep{Hart2022}. Similarly, variations exist regarding the duration of the infectious period \citep{byrne2020inferred}. The central values reported for Weibull shape parameter $k$ and the scale parameter $\lambda$ in \cite{ferretti2020} are 2.826 and 5.665, respectively. In this paper, we consider MGT varying between 4 and 8 days with an interval of one day, and  $\bar{\tau}_\gamma=14$ or $\bar{\tau}_\gamma=7$ days. Figure~\ref{fig:iprofile} shows the infectiousness profiles corresponding to different MGTs and $\bar{\tau}_\gamma$.

\begin{figure}[htbp]
\centering
\includegraphics[width=0.5\textwidth]{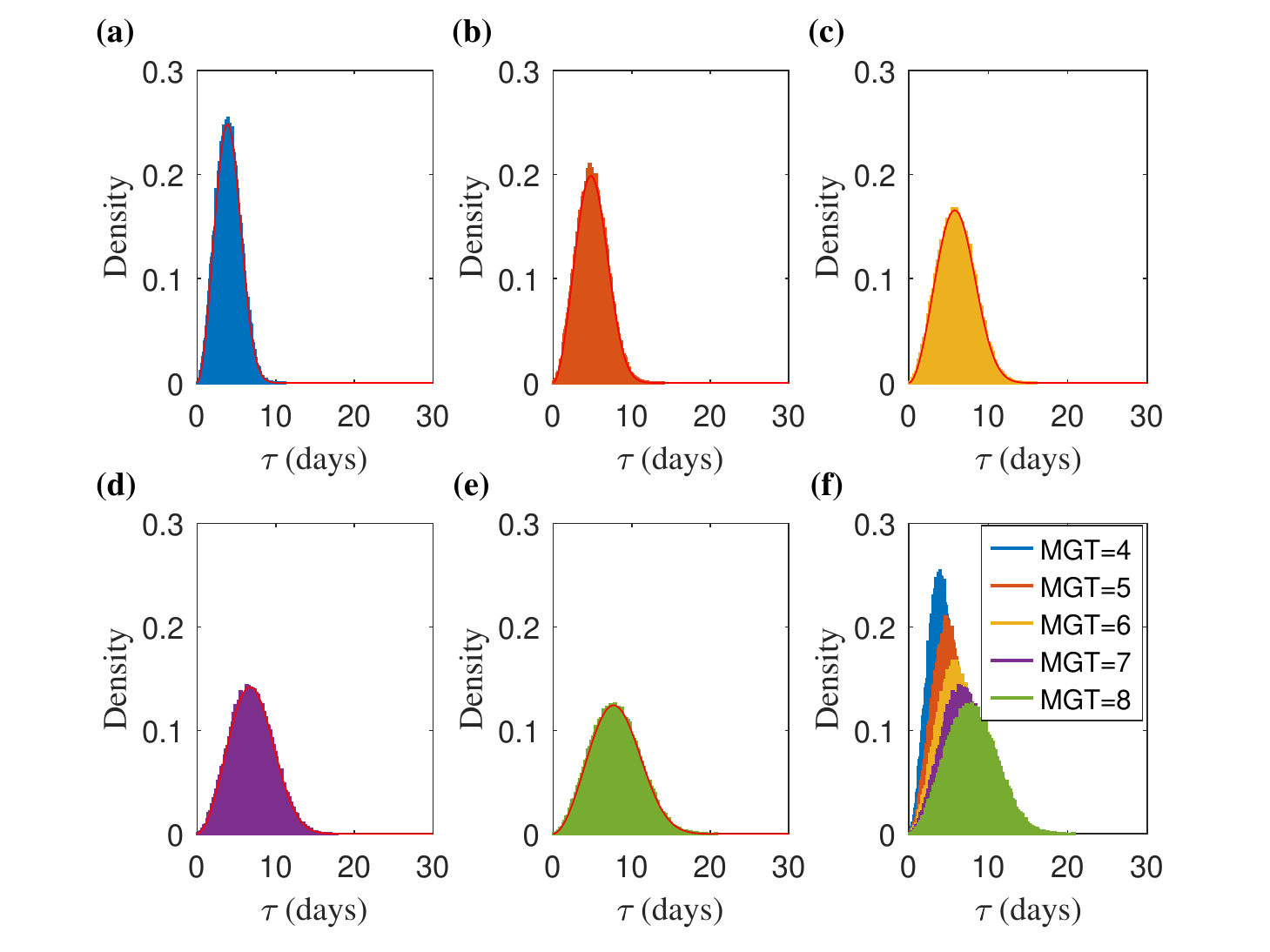}
\caption{\label{fig:gt6} Generation time distributions at the beginning of the epidemic. Colors (blue, orange, yellow, purple, green) correspond to the five generation time distributions, measured from the ABM, with mean values varying from 4 to 8 days with an interval of 1 day. In panels (a) -- (e), red curves refer to the theoretical generation time distributions for each scenario. The generation times in each scenario follow the Weibull distribution with the same shape parameter $k$ and varied scale parameter $\lambda$. In the simulations, the mean infectious period $\bar{\tau}_\gamma$ is set as 14 days, and the population size is 500,000.}
\end{figure}

In the ABM, we record the infection times between infector and infectee for all simulation runs associated with index cases and plot their distributions in Fig.~\ref{fig:gt6}. Index cases refer to those individuals infected at the beginning of the epidemic, who are used to introduce the disease into the population. The mean infectious period $\bar{\tau}_\gamma$ used to generate Fig.~\ref{fig:gt6} is equal to 14 days. However, its precise value is irrelevant to the measured Weibull generation time distribution as long as it is far enough from 0. This fact, numerically verified for several values of $\bar{\tau}_\gamma$, confirms that the computation of $\beta(\tau)$ from Eq.~\eqref{w_beta} counterbalances the recovery effects (see Fig.~2).

\subsection{Scenarios without vaccinations}
\begin{figure*}[htbp]
\centering
\includegraphics[width=\textwidth]{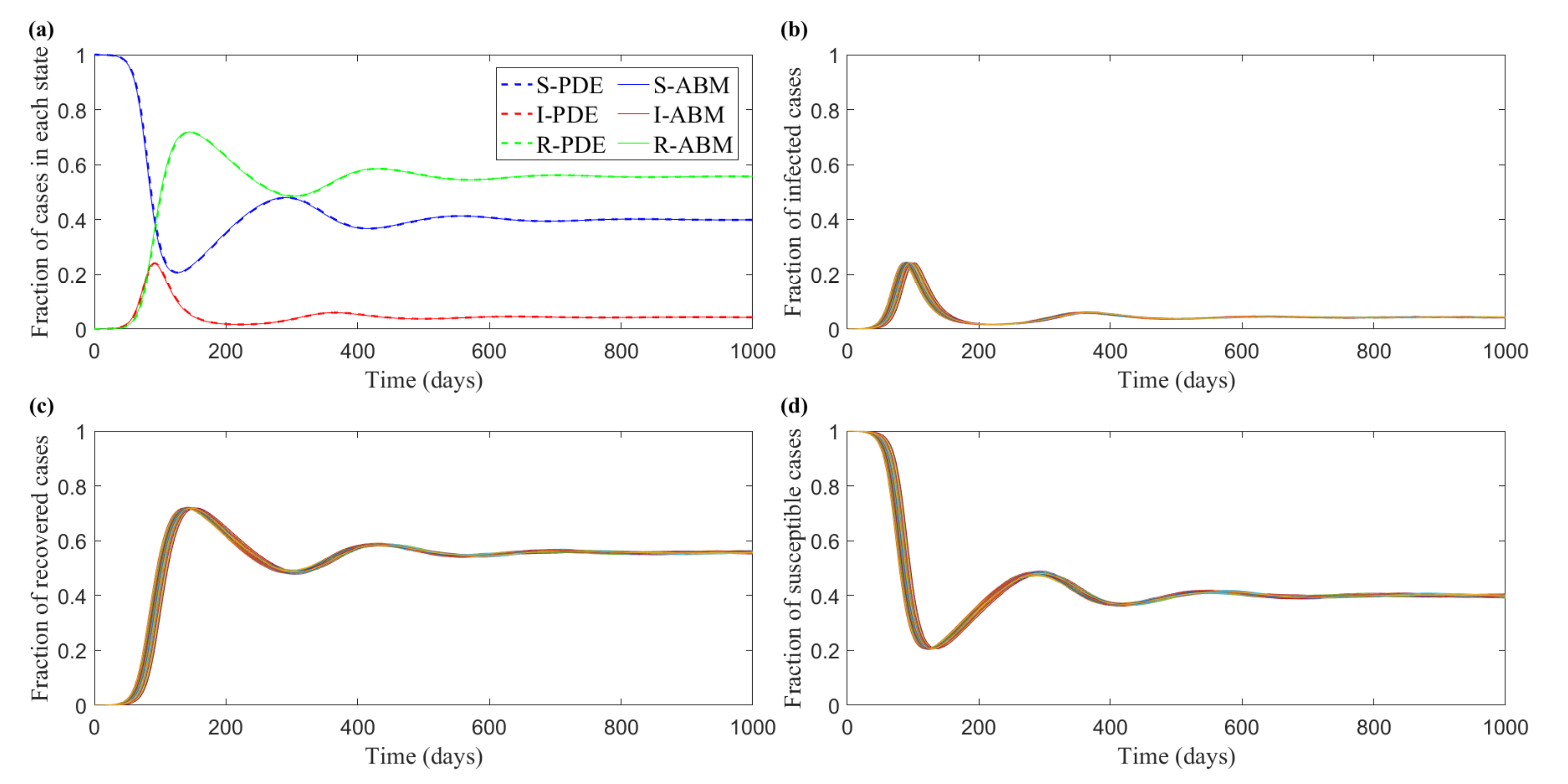}
\caption{\label{fig:baseABMConst} Simulation results with constant infectiousness profiles and a mean infectious period $\bar{\tau}_\gamma=14$ days. The dashed lines in panel (a) depict fractions of cases in each state from the numerical integration of the PDE model. The solid lines show the median value of simulation runs with secondary waves from the ABM. Panels (b) -- (d) plot the fractions of infected, recovered, and susceptible cases, respectively, for all simulation runs. Various colors represent the values resulting from different simulation runs. Five hundred simulation runs are performed with a population size of 500,000.}
\end{figure*}

\begin{figure*}[htbp]
\centering
\includegraphics[width=\textwidth]{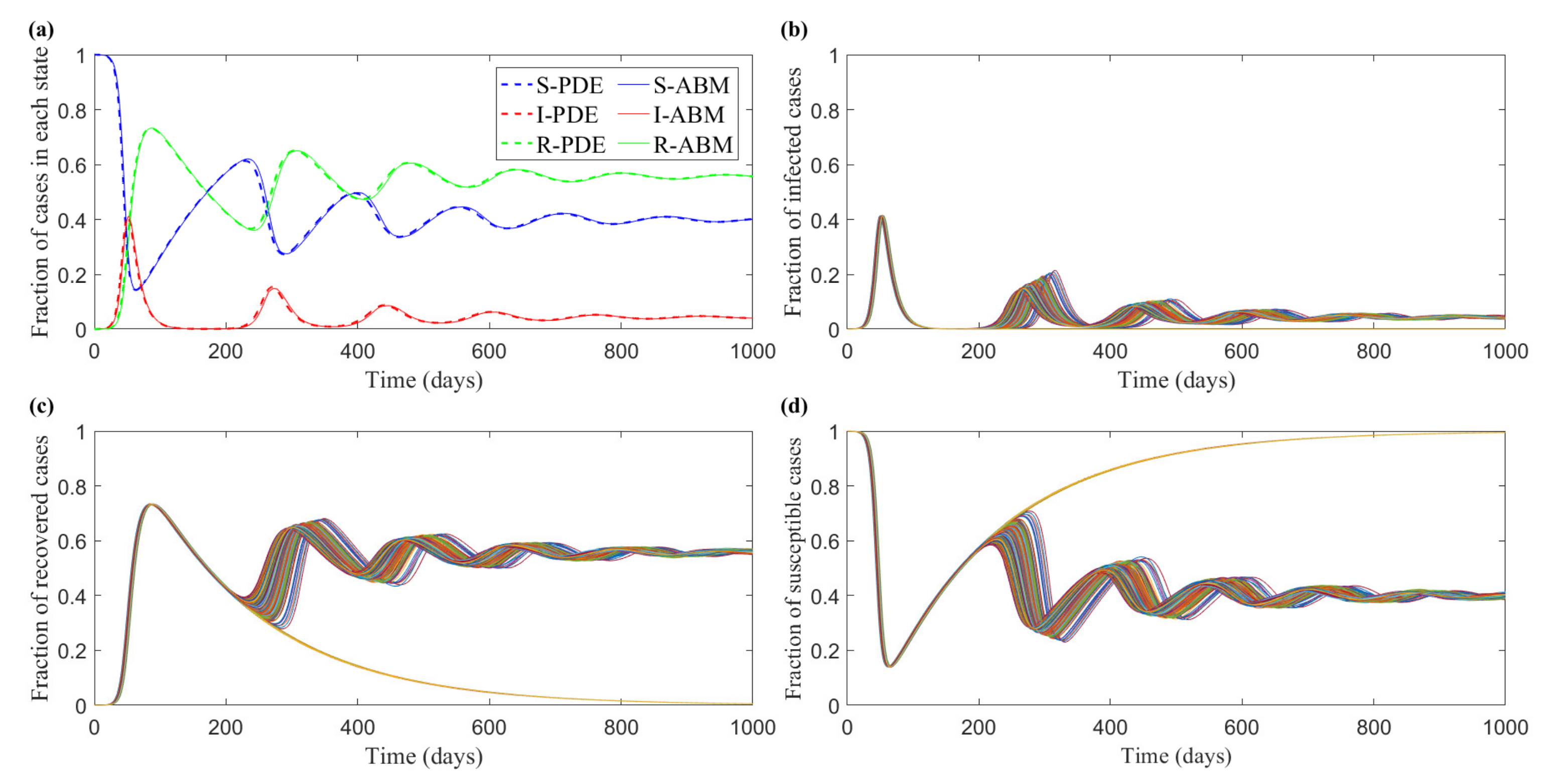}
\caption{\label{fig:baseABMWeibull} Simulation results with varying infectiousness profiles (MGT=5 days) and a mean infectious period $\bar{\tau}_\gamma=14$ days.  In panel (a), the dashed lines depict the fraction of individuals obtained from the numerical integration of the PDE model, and the solid lines show the median value of simulation runs with secondary waves from the ABM. In panels (b) -- (d), various colors represent the values resulting from different simulation runs. Panel (b) plots the fraction of infected cases for all simulation runs. Panels (c) and (d) show the fractions of recovered and susceptible cases for all simulation runs. Five hundred simulation runs are performed with a population size of 500,000.}
\end{figure*}

In this section, 0.01\% of the total population is initially infected, and the remaining population is susceptible at the beginning of the epidemic.  Without considering vaccinations, the results from ABM and the PDE model are presented in Figs.~\ref{fig:baseABMConst} and \ref{fig:baseABMWeibull}.

Overall, the model with time-varying infectiousness profiles (Fig.~\ref{fig:baseABMWeibull}) leads to more oscillations and of greater amplitude when compared to that with constant infectiousness profiles (Fig.~\ref{fig:baseABMConst}). Due to their stochastic nature, ABMs provide extra patterns not observed in the PDE model. As ABMs treat each individual as an agent, and PDE models can have fractions of an individual, all curves with PDEs are associated with secondary waves resulting from a damped oscillatory approach to endemic equilibrium. At the same time, in the ABM (Fig.~\ref{fig:baseABMWeibull}), 454 out of 500 (90.80\%) simulation runs result in secondary waves, and the rest die out after the first epidemic wave. Accordingly, we present the median values of simulation runs with secondary waves and the PDE results in Fig.~\ref{fig:baseABMConst} (a) and Fig.~\ref{fig:baseABMWeibull} (a). With a population size equaling 500,000, the ABM results resemble PDE results very well. 

\begin{figure}[htbp]
\centering
\includegraphics[width=86mm]{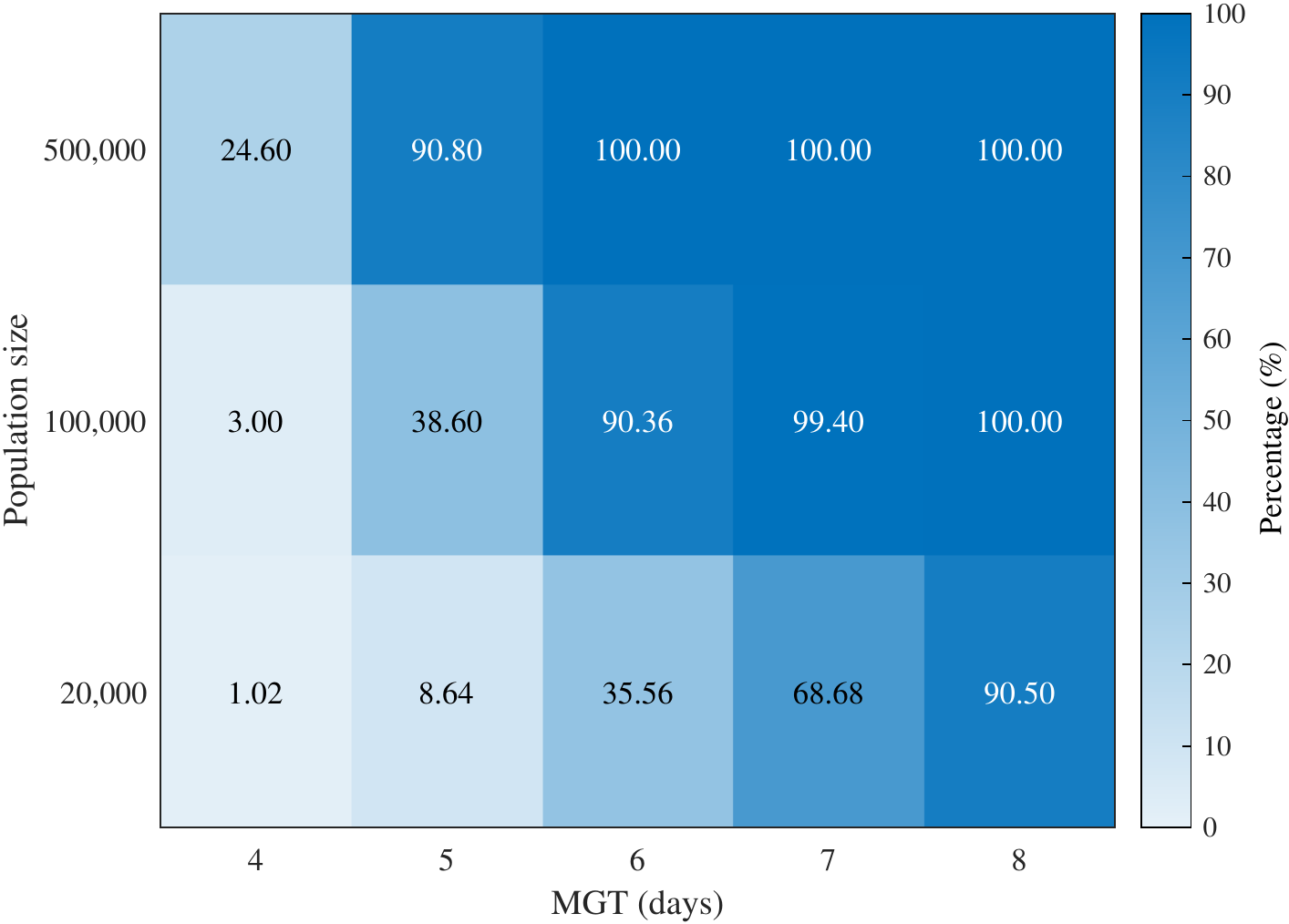}
\caption{\label{fig:risk} The impact of population size and the mean generation time on the risks for secondary wave occurrences. In the figure, MGT stands for the mean generation time, which varies from 4 to 8 days with an interval of 1 day. The mean infectious period $\bar{\tau}_\gamma=14$ days. The value in the heatmap indicates the percentage of simulation runs with secondary epidemic waves, given a first epidemic peak. Simulation runs associated with initial extinctions or without selected index cases are not considered. In all these cases, the predicted endemic equilibrium is eventually reached.}
\end{figure}

Figure~\ref{fig:risk} shows the impact of the population size and infectiousness profiles on the risk for secondary waves. We can see that the chance for secondary wave occurrences rises along with the increase in MGT. Probably, this is due to the fact that those individuals who remain infectious for a long time have higher infectiousness towards the end of their epidemic period as the MGT increases because the infectiousness profile gets stretched out to the right. On the other hand, the percentage of simulation runs with secondary epidemic waves in populations of size 20,000 is lower than that in populations of size 100,000 and 500,000. So, with the same fraction of infected cases, larger populations  have a higher chance for secondary outbreaks over small populations. 

Simulation runs without secondary waves may be associated with epidemics dying out after the first peak, with initial extinctions, or with no index cases. Focusing on whether the secondary waves appear after the first peak, we calculate the risk for secondary waves in Fig.~\ref{fig:risk}, without taking into account those simulation runs associated with initial (stochastic) extinctions or those without selected index cases. At the beginning of each simulation run, each agent enters the infected state based on a probability of 0.0001. This leads to stochasticity in the number of index cases and thus to simulation runs that fail to introduce index cases with populations of size 20,000, for which the expected number of index cases is only 2. Let us denote the percentages of total simulation runs with initial extinctions and without index cases as $p_{ex}$ and $p_{no}$, respectively. The values of $p_{ex}$ and $p_{no}$ become substantially small as the population size increases. For example, with populations of size 20,000, $p_{ex}=10.40\%$ ($p_{no}=11.00\%$) for MGT=4 days, and $p_{ex}=15.80\%$ ($p_{no}=12.60\%$) for MGT=8 days. With populations of size 100,000, $p_{ex}=0.4\%$ ($p_{no}=0$) for MGT=8 days, and, with populations of size 500,000, $p_{ex}=0$ ($p_{no}=0$) for MGT=8 days.

\begin{figure}[htbp]
\centering
\includegraphics[width=0.5\textwidth]{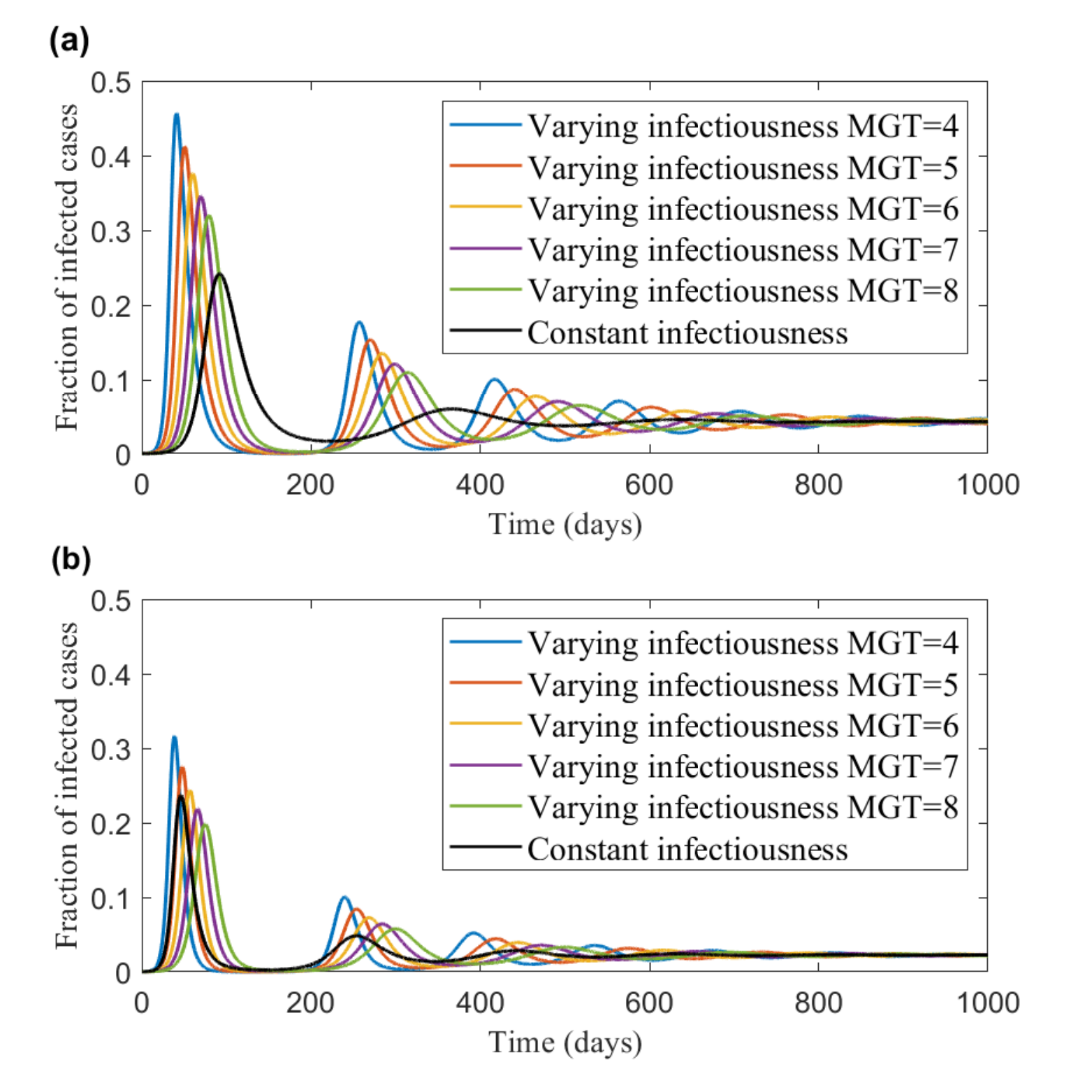}
\caption{\label{fig:ep1} Impact of infectiousness profiles on the fraction of infected individuals. In panel (a), the infectious period $\bar{\tau}_\gamma=14$ days. In panel (b), $\bar{\tau}_\gamma=7$ days. A shorter infectious period (faster recovery) interferes more with the generation time distribution (transmission process) and reduces the differences among curves. Figures in both panels are based on the numerical integration of the PDE model.}
\end{figure}

Figure~\ref{fig:ep1} depicts the fractions of infected individuals obtained by considering variations in infectiousness profiles. These fractions follow from the numerical integration of the PDE model and are given by $\int_0^\infty I(t,\tau) \, d\tau / N$. In Fig.~\ref{fig:ep1} (a), with the mean infectious period $\bar{\tau}_\gamma=14$ days, when the MGT increases from 4 to 8 days, the peak time for the epidemic waves is postponed (from day 41 to day 79 for the first peak time) with height reduced by 29.95\%. This shift in the time of the first peak is consistent with lower initial epidemic growth rates predicted by Eq.~\eqref{Euler_Lotka} for larger MGTs. Precisely, the predicted values of the initial growth rate are as follows: $r$(MGT=4)$=0.2464$, $r$(MGT=5)$=0.1971$, $r$(MGT=6)$=0.1642$, $r$(MGT=7)$=0.1408$, $r$(MGT=8)$=0.1235$, and $r=0.1071$ for constant $\beta$. All of them are in agreement with the initial growth rates estimated from Eq.~\eqref{Euler_Lotka}. In comparison, differences between epidemic curves due to the shift in the infectiousness profiles are less pronounced when the range of MGT gets closer to the mean infectious period, e.g., $\bar{\tau}_\gamma=7$ in Fig \ref{fig:ep1} (b). Remarkably, the only initial growth rate that has changed is the one for constant $\beta$, which, now, is higher than those corresponding to the largest MGT (predicted initial growth rate for constant $\beta$: $r=0.2143$). This is due to the fact that, for constant $\beta$, the generation time distribution is equal to the distribution of the length of the infectious period (cf. with Eq.~\eqref{w_beta}). The rest of the generation time distributions are independent of the recovery rate and, so, the corresponding curves are arranged in the same order in both panels. This indicates that the differences between models based on time-varying and constant infectiousness profiles heavily depend on the recovery processes. When recovery processes interfere less with infection processes, we can see noticeable effects of infectiousness profiles. 

\subsection{Scenarios with vaccinations}
\begin{figure*}[htbp]
\centering
\includegraphics[width=\textwidth]{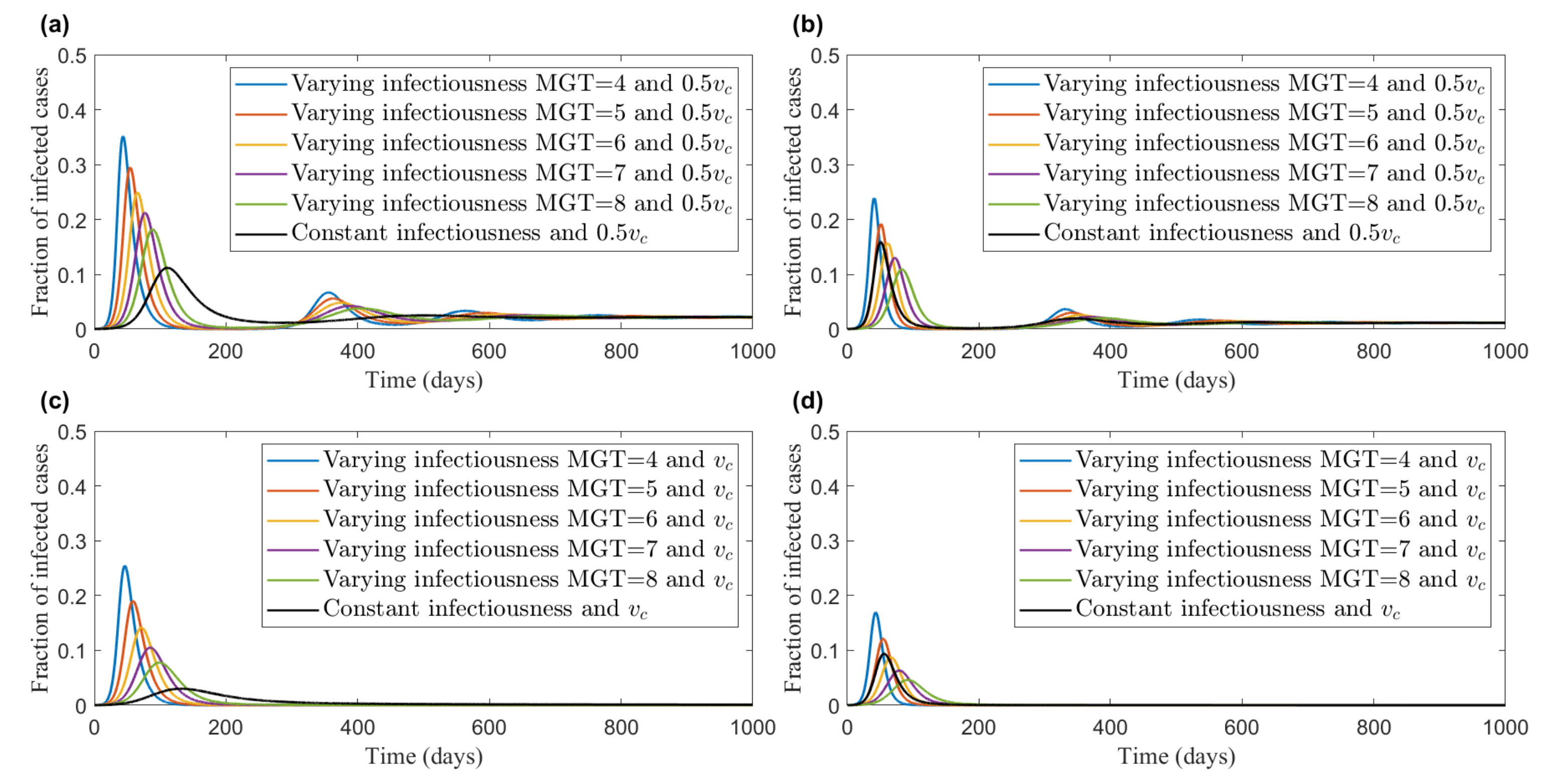}  
\caption{\label{fig:fig89} Scenarios with vaccinations. All individuals are initially susceptible except 0.01\% of the total population which is set as index cases. In panels (a) and (c), the mean infectious period $\bar{\tau}_\gamma=14$ days, and the vaccination rate equals $0.5v_c$ and $v_c$, respectively. In panels (b) and (d), $\bar{\tau}_\gamma=7$ days and the vaccination rate equals $0.5v_c$ and $v_c$, respectively. Figures are obtained from the numerical integration of the PDE model.}
\end{figure*}

Figure~\ref{fig:fig89} shows the fractions of infected individuals, computed again as $\int_0^\infty I(t,\tau)\,d\tau/N$, with a uniform vaccination rate $v$ equal to $0.5v_c$ and $v_c$, and different infectiousness profiles. Here, the critical vaccination rate $v_c = ({\mathcal{R}_0-1})/\overline{\tau}_{\delta^v}$ equals $0.00825$. All individuals are initially susceptible except 0.01\% of the total population which is set as index cases to start the epidemic. We can see that the infectiousness profiles and the mean infectious period affect the epidemic dynamics in the transient phases but have no impact on the critical vaccination as long as the parameters can achieve the same $\mathcal{R}_0$.    

\begin{figure*}[htbp]
\centering
\includegraphics[width=\textwidth]{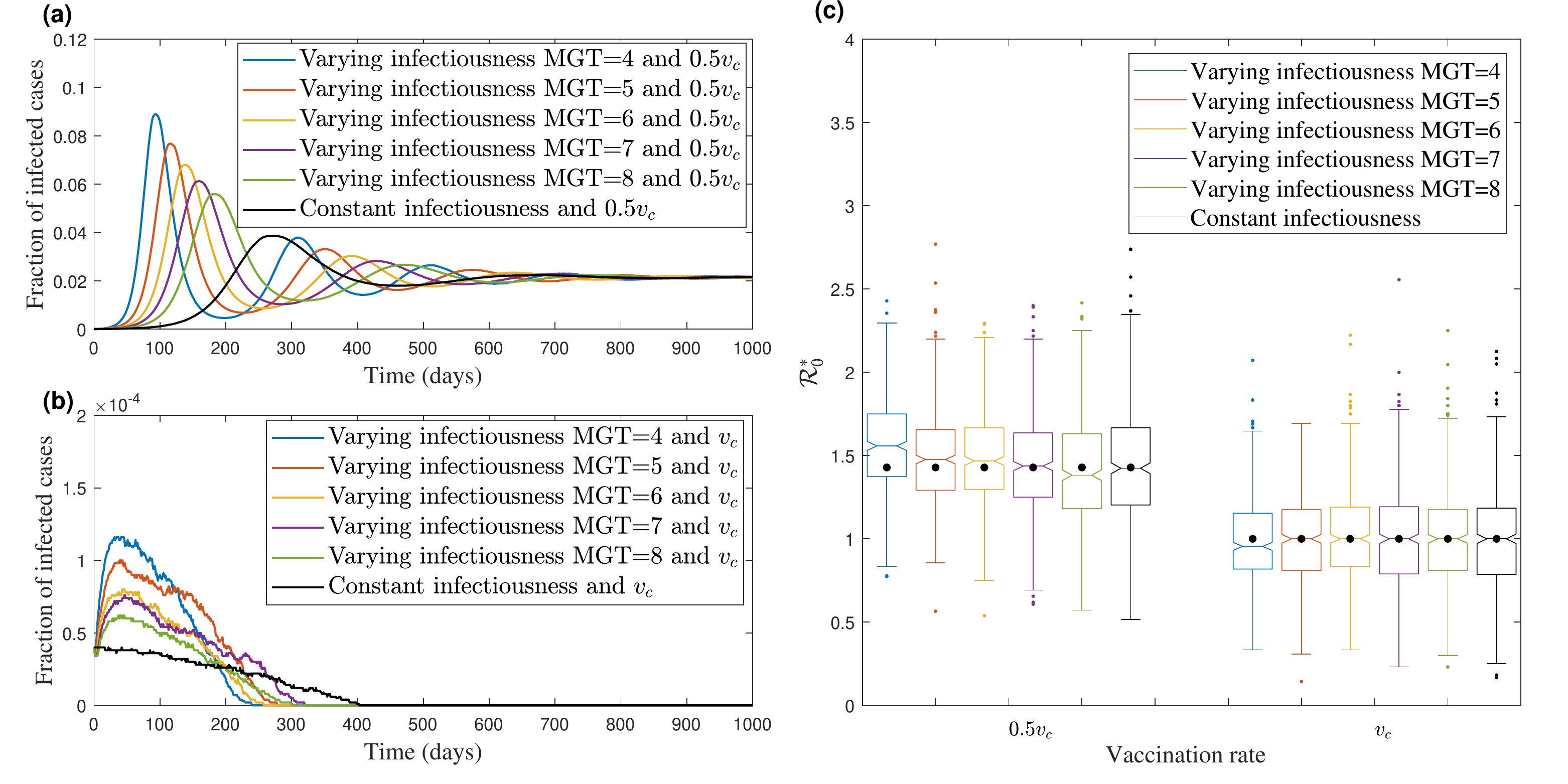}
\caption{\label{fig:fig1011} Scenarios with initially vaccinated individuals. The vaccinated population is initially present at a fraction given by the DFE, and 0.01\% of the susceptible population is initially infected. Panels (a) – (b) plot the median fractions of infected cases from ABM. The vaccination rate equals $0.5v_c$ and $v_c$ in panels (a) and (b), respectively. Panel (c) depicts boxplots of the measured $\mathcal{R}_0^*$ through ABM and the black solid circles reflect the predicted $\mathcal{R}_0^*$. $\mathcal{R}_0^*$ denotes the basic reproduction number at the DFE considering vaccinations. The mean infectious period $\bar{\tau}_\gamma$ is 14 days. Five hundred simulation runs are performed for each scenario with a population size of 500,000.}
\end{figure*}

On the other hand, we consider scenarios  {(Fig.~\ref{fig:fig1011})} where the vaccinated population is present at the beginning of the epidemic, and 0.01\% of the susceptible population is initially infected. More specifically, the fraction of the susceptible population is obtained through the DFE condition $S^*/N = {1}/({1+v\,\overline{\tau}_{\delta^v}})$, and the rest of the population is vaccinated ($\int_0^\infty V^*(\tau)\,d\tau/N =1-S^*/N$). Accordingly, we have 40\% and 57.14\% of the total population susceptible at the beginning of the epidemic for scenarios with $v=v_c$ and $v=0.5v_c$, respectively. 

Figure~\ref{fig:fig1011} shows the evolution of the  median fractions of infected cases resulting from simulations where vaccinated people are initially present. As expected, when new cases are introduced, outbreaks are contained very well under scenarios with $v=v_c$, with only a very small fraction of infections. In comparison, there are large outbreaks under scenarios with $v=0.5\,v_c$. For example, with MGT=4 days, the peak fraction of infected cases in Fig.~\ref{fig:fig1011} (a) is 0.0889 while the peak fraction of infected cases in Fig \ref{fig:fig1011} (b) is $1.16\times 10^{-4}$. Considering a population size of 500,000, a peak fraction equal to $1.16\times 10^{-4}$ means that only 58 individuals are infected at peak time. So, at the critical vaccination rate $v=v_c$, the introduction of new infections at the start of the simulation leads to only minor outbreaks. Fig.~\ref{fig:fig1011} (a) also shows that the initial growth rates are lower than without vaccination, but are ordered in the same way. As additional information, Fig.~\ref{fig:fig1011} (c) depicts the measured $\mathcal{R}_0^*$ through simulations, which is the number of secondary cases divided by the number of index cases. Since the initial fractions of vaccinated and susceptible individuals are given by the DFE, $\mathcal{R}_0^*$ can be interpreted as the basic reproduction number at the DFE considering vaccinations. At the critical vaccination rate $v_c$, the mean value for $\mathcal{R}_0^*$ ranges from 0.9895 to 1.0182. At $v=0.5\,v_c$, the mean value for $\mathcal{R}_0^*$ varies between 1.4069 and 1.5645. Under scenarios of $v=v_c$ and $v=0.5\,v_c$, the predicted values for $\mathcal{R}^*_0$, namely, $\mathcal{R}_0 S^*/N$, are 1 and 1.4285, respectively, which are within the observed ranges. 

\section{DISCUSSION}
This paper presents a general SIRVS model considering waning immunity and age of infection. We analyzed how variations in infectiousness profiles under the same $\mathcal{R}_0$ could affect the epidemic dynamics. Compared with Markovian models, non-Markovian models with time-varying infectiousness profiles create more damped oscillations with peak times affected in the transient phases. Remarkably, the magnitude of this difference between the two types of models heavily depends on the recovery processes. When the recovery process interferes more with the infection processes, the variations between models become less pronounced. Such an interference is possible because, in the standard formulation of epidemic models with age of infection (see, for instance, \cite{Nakata}), recovery and infectiousness are modeled as independent of each other. This modeling assumption, however, is clearly questionable if infectiousness is interpreted in terms of viral load and recovery occurs only once a low level viral load is reached.

We have also seen that different combinations of infectiousness profiles and infectious periods have no impact on the critical vaccination $v_c$ as long as they lead to the same $\mathcal{R}_0$. Indeed, given $\mathcal{R}_0$, the mean duration of the recovery period is the only feature of its profile that determines the value of $v_c$. However, when vaccination rates are lower than the critical rate $v_c$, models with time-varying infectiousness still have a transient behavior with damped oscillations of higher amplitude than Markovian models and retain the same order of the initial growth rates. This echoes the findings of \cite{favero2022modelling}, which found that vaccination reduces the reproduction number without changing the generation time distribution during the epidemic. Besides, with susceptible and vaccinated people at the beginning of the epidemic, a population at the predicted critical vaccination rate is resilient to future epidemics, regardless of the particular infectiousness profile.

Loss of immunity is one of the causes of the oscillations observed in epidemic models. For instance, if there is a constant period of temporary immunity, destabilization of the endemic equilibrium of the SIRS model is possible through a Hopf bifurcation (\cite{Hethcote}). As for damped oscillations, they occur in the standard (Markovian) SIRS model, and an approximation of their period is also well known (\cite{Keeling}). Here we have explored the impact of the infectiousness profile on the occurrence and shape of these oscillations.

We have found that ABMs not only can produce results close to the PDE formulation with large population sizes, but also provide additional insights into the risk of secondary waves that are not obtained under the latter formulation. They suggest that, even with large populations, epidemics could die out after an initial epidemic peak if the decline in prevalence is fast enough. The occurrence of these waves then depends on both population size and infectiousness profile (through the assumed mean generation time). Moreover, since they are always associated with an endemic equilibrium, if stochastic extinction after the first peak is avoided, the convergence towards endemic equilibrium always occurs because the damped behavior of the oscillations prevents a return to very low levels of prevalence. Besides, at the same population size, the percentages of simulations with secondary waves with constant infectiousness are higher than those with varying infectiousness. Therefore, given the importance of reducing the risks of the emergence of secondary waves during the course of an epidemic, it highlights the importance of selecting the appropriate modeling approach and estimating the generation time distributions to tackle future epidemics.

\section*{Acknowledgement}
The work of Q.Y. and C.S. has been supported by the National Science Foundation under grant award no. CMMI-1744812. J.S. has been partially supported by Grant No. PID2019-104437GB-I00 of the Agencia Estatal de Investigación, Ministerio de Ciencia e Innovación of the Spanish government, and is a member of the Consolidated Research Group 2021 SGR 00113 of the Generalitat de Catalunya. Any opinions, findings, and conclusions, or recommendations expressed in this material are those of the authors and do not necessarily reflect the views of the funding agencies.

\appendix*
\section{Simulation results with non-Markovian recovery processes}
Denote $\gamma(\tau)$ as the age-dependent recovery rate, and $\psi_{ip}(\tau)$ as the infectious period distribution. This distribution can be characterized as recovery processes, which can be expressed as follows:  
\begin{equation}
\psi_{ip}(\tau) = \gamma(\tau)e^{-\int_{0}^{\tau} \gamma(s) \, ds}.
\label{psi_ip}
\end{equation}

In the following, we consider infectious periods (from I to R compartments) following the Weibull distribution, $\psi_{ip}(\tau) =\frac{\alpha}{\mu}(\frac{\tau}{\mu})^{\alpha-1}e^{-(\tau/\mu)^\alpha}$, where $\alpha$ is the shape parameter and $\mu$ is the scale parameter of the infectious period distribution. According to Eq.~\eqref{psi_ip}, we have $\gamma(\tau) = \frac{\alpha}{\mu} \left(\frac{\tau}{\mu}\right)^{\alpha-1}$.

From Eq~\eqref{w_beta}, 
it follows that 
\begin{equation*}
\beta(\tau) = \frac{w(\tau) R_0}{c \, e^{-\int_{0}^{\tau} \gamma(s) \, ds}}.
\label{beta_tao1}
\end{equation*}

\noindent
Since now $\gamma(\tau) = \frac{\alpha}{\mu} \left(\frac{\tau}{\mu}\right)^{\alpha-1}$, we have $\beta(\tau) = w(\tau) R_0/\left(c \, e^{-\left(\frac{\tau}{\mu}\right)^\alpha}\right)$. 
In the simulations, we set $\alpha = 2.82$ and $\mu = 15.72$ to obtain the mean infectious period $\overline{\tau}_\gamma = 14$ days. The parameters of $w(\tau)$ are the same as those in the main text.

Without considering vaccinations, the results from agent-based simulations are plotted in Fig. A.1. We record the recovery times for all simulation runs associated with index cases, from which we compute the corresponding infectious periods and plot their distribution in Fig. A.1(d). In Fig.~\ref{fig:a1}, 493 out of 500 (98.6\%) simulation runs result in secondary waves, while the rest dies out after the first epidemic wave. In comparison, with the same MGT (5 days) and a constant recovery rate equal to the inverse of a mean infectious period $\overline{\tau}_\gamma = 14$ days, 90.8\% simulation runs result in secondary waves. It suggests that, with the same mean infectious period, the percentage of secondary wave occurrences increases when the recovery rate changes from constant to nonconstant values.

In Fig.~\ref{fig:a2}, we present the median values of simulation runs with secondary waves. Consistent with the patterns observed in Fig. 7, we can observe that as the MGT increases from 4 to 8 days, the peak time for the epidemic waves is postponed (from day 42 to day 79 for the first peak time) with a reduced height of 35.91\%. When comparing the results with constant recovery rate shown in Fig. 7, we notice that the disease prevalence increases when infectious periods are Weibull distributed but with the same $\overline{\tau}_\gamma = 14$ days. 

\newcounter{figureminus9}
\setcounter{figureminus9}{0}

\renewcommand{\thefigure}{A.\arabic{figureminus9}}
\setcounter{figureminus9}{\value{figure}}
\addtocounter{figureminus9}{-9}
\addtocounter{figureminus9}{1}

\begin{figure*}[htbp]
\centering
\includegraphics[width=\textwidth]{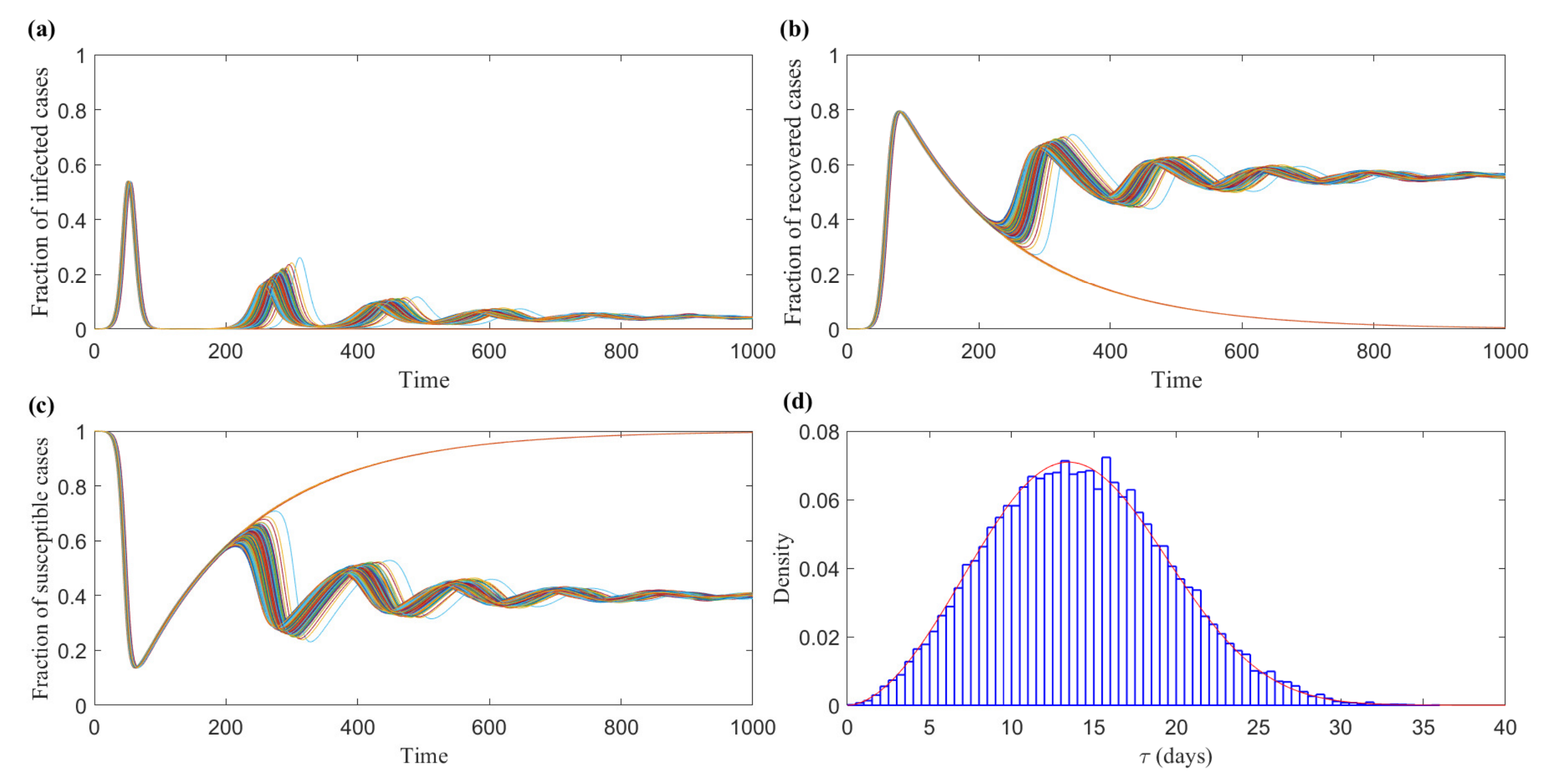}  
\caption{Simulation results with varying infectious profiles (MGT=5 days) and non-Markovian recovery processes. The Weibull infectious period distribution is associated with shape parameter $\alpha=2.82$ and scale parameter $\mu=15.72$, leading to a mean infectious period $\overline{\tau}_\gamma = 14$ days. Panels (a) – (c) plot the fractions of infected, recovered, and susceptible cases, respectively, for all simulation runs. Various colors represent the values resulting from different simulation runs. In panel (d), we present the infectious period distribution measured from ABM, and the red curve refers to the theoretical infectious period distribution. Here 0.01\% of the susceptible population is initially infected. Five hundred simulation runs are performed with a population size of 500,000.
}
\label{fig:a1}
\end{figure*}

\addtocounter{figureminus9}{1}
\begin{figure}[htbp]
\centering
\includegraphics[width=0.5\textwidth]{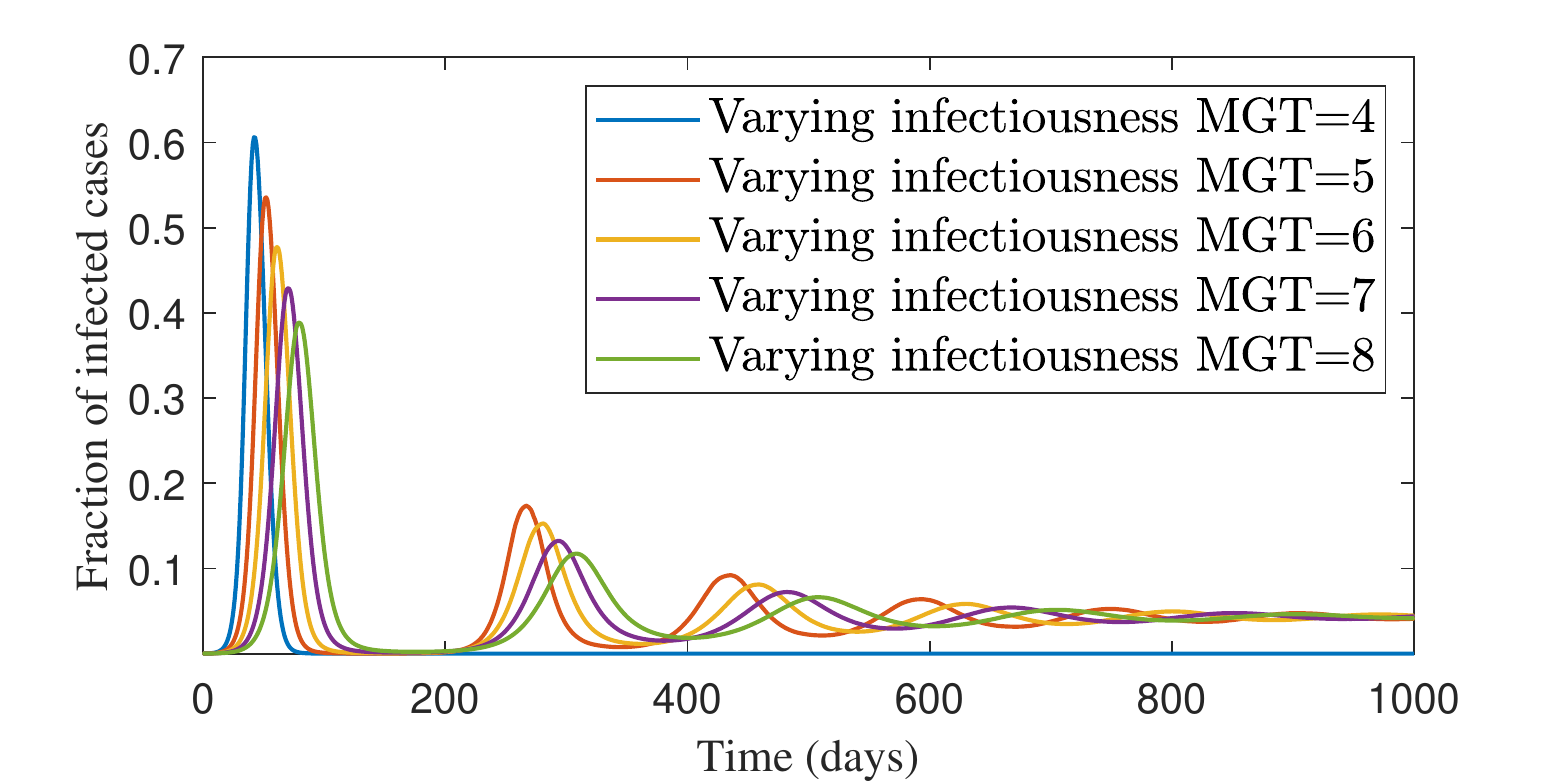}
\caption{Scenarios with different infectiousness profiles and the same non-Markovian recovery time distributions. The Weibull infectious period distribution is associated with a shape parameter $\alpha=2.82$ and a scale parameter $\mu=15.72$, resulting in a mean infectious period $\overline{\tau}_\gamma = 14$ days. The figure depicts the median fraction of infected cases from the ABM. Initially, 0.01\% of the susceptible population is infected. Five hundred simulation runs are performed for each scenario with a population size of 500,000.
}
\label{fig:a2}
\end{figure}

\end{document}